\pdfoutput=1
\documentclass[usenatbib]{mnras}
\usepackage{graphicx,times}

\newcommand{\HI}{\mathrm{H\,I}}

\newcommand{\HeI}{\mathrm{He\,I}}
\newcommand{\HeII}{\mathrm{He\,II}}

\newcommand{\HIa}{H\,{\sevensize{{I}}}\,\,}
\newcommand{\HIb}{H\,{\sevensize{{I}}}}

\newcommand{\HeIIa}{He\,{\sevensize{{II}}}\,\,}
\newcommand{\HeIIb}{He\,{\sevensize{{II}}}}

\newcommand{\GHI}{\Gamma_{\HI}}
\newcommand{\GHeII}{\Gamma_{\HeII}}
\newcommand{\etath}{\eta_{\mathrm{thin}}}

\newcommand{\lya}{Ly$\alpha$ }

\newcommand{\teff}{\tau_\mathrm{eff}}

\title[Self-consistent helium-ionizing background]{A self-consistent 3D model of fluctuations in the helium-ionizing background}
\author[F.~B. Davies, S.~R. Furlanetto, K.~L. Dixon]{Frederick B. Davies$^{1,2}$\thanks{davies@mpia.de}, Steven R. Furlanetto$^2$, Keri L. Dixon$^3$\\
$^1$Max-Planck-Institut f{\"u}r Astronomie, K{\"o}nigstuhl 17, D-69117 Heidelberg, Germany \\
$^2$Department of Physics \& Astronomy, University of California, Los Angeles, Box 951547, Los Angeles, CA 90095 \\
$^3$Astronomy Centre, University of Sussex, Falmer, Brighton BN1 9QH, UK}

\begin{document}

\maketitle

\begin{abstract}
Large variations in the effective optical depth of the \HeIIa \lya forest have been observed at $z\ga2.7$, but the physical nature of these variations is uncertain: either the Universe is still undergoing the process of \HeIIa reionization, or the Universe is highly ionized but the \HeIIb-ionizing background fluctuates significantly on large scales. In an effort to build upon our understanding of the latter scenario, we present a novel model for the evolution of ionizing background fluctuations. Previous models have assumed the mean free path of ionizing photons to be spatially uniform, ignoring the dependence of that scale on the local ionization state of the intergalactic medium (IGM). This assumption is reasonable when the mean free path is large compared to the average distance between the primary sources of \HeIIb-ionizing photons, $\ga L_\star$ quasars. However, when this is no longer the case, the background fluctuations become more severe, and an accurate description of the average propagation of ionizing photons through the IGM requires additionally accounting for the fluctuations in opacity. We demonstrate the importance of this effect by constructing 3D semi-analytic models of the helium ionizing background from $z=2.5$--$3.5$ that explicitly include a spatially varying mean free path of ionizing photons. The resulting distribution of effective optical depths at large scales in the \HeIIa \lya forest is very similar to the latest observations with HST/COS at $2.5 \la z \la 3.5$.
\end{abstract}


\section{Introduction}
The hydrogen and helium reionization epochs produce some of the largest scale features in the Universe. They reflect the cumulative ionizing photon output of galaxies and black holes, or in other words, the history of star formation and supermassive black hole accretion across cosmic time. Helium reionization is believed to finish at $z\sim3$--$4$, when quasars start to dominate the ionizing photon production in the universe (e.g. \citealt{Sokasian2002,WL2003,FO2008}). Occurring near the peak era of star and black hole formation, this is the last global transition experienced by the intergalactic medium (IGM) and significantly heats that material (e.g. \citealt{Theuns2002,HH2003,Bolton2009}). This heating indirectly affects the star formation history of the Universe (by influencing its fuel supply) as well as observables like the \HIa \lya forest. The \HeIIb-ionizing background also offers the prospect of shedding light on the sources that drive it, including the abundance, spectra, lifetimes, and emission geometries of quasars (e.g. \citealt{WW2006,Worseck2007,FL2011}).

Observations of substantial transmission through the \HeIIa \lya forest below $z\sim2.7$ \citep{Davidsen1996,Kriss2001,Zheng2004,Fechner2006} suggest that helium reionization is complete by that epoch. At $z \ga 2.7$, however, the \HeIIa \lya forest opacity rapidly increases and significant fluctuations in the \HeIIa \lya effective optical depth $\tau_\mathrm{eff}$ arise on scales ranging from $10$--$100$ comoving Mpc (e.g. \citealt{Reimers1997,Heap2000,Shull2004}). Recent observations with HST/COS have confirmed these features \citep{Shull2010,Worseck2011,SS2013,SS2014} and continue to increase the number of \HeIIa \lya forest sightlines \citep{Syphers2012,Worseck2014a}. Straightforward models of the \HeIIa ionizing background with a uniform mean free path of ionizing photons cannot explain the fluctuations observed at $z\ga2.7$ \citep{FD2010}, suggesting that the culprit is incomplete helium reionization (see also \citealt{Worseck2011}). Interestingly, the \HIa \lya forest above $z\sim5.5$ shows a similar increase in $\tau_\mathrm{eff}$ fluctuations after the epoch of hydrogen reionization has finished \citep{Fan2006,Becker2015}, which may be due to a spatially varying mean free path (\citealt{DF2015}, but see \citealt{DAloisio2015}). The incomplete reionization interpretation also appears to be in moderate tension with the existence of regions with significant \HeIIa transmission at $z\sim3.5$ \citep{Worseck2014a,Compostella2014}. Unfortunately, the era immediately following \HeIIa reionization has not been well-studied, as radiative transfer simulations (which have mostly focused on the reionization process) are expensive and hence limited to ``best-guess" parameters. 

\citet[][henceforth DF14]{DF2014} showed that fluctuations in the ionizing background could play an important role in the evolution of the \emph{mean} background by inducing spatial variations in the opacity to ionizing photons. Toward higher redshifts, as the opacity of the IGM to \HeIIb-ionizing photons increases, the fluctuations in the background similarly increase \citep{Furlanetto2009}, leading to a rapidly evolving ionizing background. However, once these fluctuations become an important source of additional opacity the assumption of uniform emission and absorption of ionizing photons in commonly used 1D cosmological radiative transfer models \citep{HM1996,HM2012,FG2009} breaks down. A full treatment of these effects requires a 3D realization of ionizing background-dependent opacity and discrete ionizing sources. In this work we construct such a model and find that it can describe the observed evolution of $\tau_\mathrm{eff}$ and its variations between sightlines decently well even under the assumption of a post-reionization universe. 

The structure of the paper is as follows. In Section~\ref{sec:opacity}, we outline our prescription for the fluctuating opacity of the IGM to \HeIIb-ionizing photons. In Section~\ref{sec:methods}, we describe our numerical model for computing the ionizing background and qualitatively discuss the effect of a fluctuating mean free path. In Section~\ref{sec:forest}, we employ a simple prescription for the \HeIIa \lya forest to relate our ionizing background model to observations. In Section~\ref{sec:uncertain}, we discuss the uncertainties of our simplified approach. Finally, in Section~\ref{sec:conclusion} we conclude by discussing the implications of our model and future improvements that will enhance its predictive power.

In this work we assume a standard $\Lambda$CDM cosmology with $H_0 = 70$ km/s/Mpc, $\Omega_m = 0.3$, and $\Omega_\Lambda = 0.7$. Distance units should be assumed to be comoving unless otherwise specified.

\section{The Opacity of the IGM to Ionizing Photons}\label{sec:opacity}

The opacity of the IGM to ionizing photons under the assumption of Poisson-distributed absorbing clouds can be parameterized by the optical depth per unit redshift,
\begin{equation}
	 	{{d\bar{\tau}}\over dz}(\nu,z) = \int_{0}^{\infty} dN_{\HI}  f(N_\HI,z) (1-e^{-\tau_\nu}),
\end{equation}
where $f(N_\HI,z)$ is the \HIa column density distribution function (CDDF), and 
\begin{equation}
\tau_\nu = N_\HI \sigma_\HI(\nu) + N_\HeI \sigma_\HeI(\nu) + N_\HeII \sigma_\HeII(\nu)
\end{equation}
is the optical depth of an absorber with \HIa column density $N_\HI$ at frequency $\nu$. We adopt the shape and normalization of the $z\approx2.5$ CDDF from \citet{Prochaska2014} and assume that the normalization evolves as $(1+z)^{2.5}$ to roughly match the measured redshift evolution of the \HIa effective optical depth \citep{Becker2013}. For \HeIIb-ionizing photons with $\nu > 4$\,$\nu_\HI = \nu_\HeII$, the opacity of an individual absorber is dominated by its \HeIIa content, which depends on the ratio of the \HIa and \HeIIa ionization rates ($\GHI$, $\GHeII$). In the optically thin limit, $N_\HeII$ can be related to $N_\HI$ by the parameter $\etath$,
\begin{equation}
\etath \equiv \frac{N_\HeII}{N_\HI} = \frac{\GHI}{\GHeII}\frac{\alpha_\HeII^A}{\alpha_\HI^A}\frac{Y}{4X} ,
\end{equation}
where $\alpha_\HI^A$ and $\alpha_\HeII^A$ are the case-A recombination coefficients of \HIa and \HeIIa, and $X$ and $Y$ are the hydrogen and helium mass fractions, respectively. For optically thick absorbers, the relationship is more complicated due to self-shielding of \HeIIa and \HIa ionizing photons. Modern cosmological radiative transfer models (e.g. \citealt{FG2009,HM2012}) determine $\eta(N_\HI)$ by computing simplified radiative transfer of the ionizing background assuming a slab geometry and Jeans ansatz for absorbers \citep{Schaye2001}. Because of the implied dependence of absorber properties on $\GHeII$, ionizing background calculations must iterate several times to achieve self-consistency between the radiation field and the \HeIIa absorber distribution. The link between $\GHeII$ and absorbers leads to an enhanced sensitivity of $\GHeII$ to evolution in the emissivity of ionizing photons (DF14; see also \citealt{McQuinn2011}).

In DF14, we extended this idea to link the \emph{local} opacity of the IGM to the \emph{local} $\GHeII$, which fluctuates significantly due to the rarity of the dominant sources of \HeIIb-ionizing photons \citep{Furlanetto2009}. In effect, this means that the mean free path of ionizing photons $\lambda$ fluctuates along with the intensity of the ionizing background. We showed that if ionizing photons sample the distribution of ionization rates $f(\Gamma)$, the overall opacity of the IGM increases due to the skewed nature of the distribution. However, naive application of this effect to a one-dimensional cosmological radiative transfer model caused the ionizing background to vanish at all redshifts unless a somewhat ad-hoc correction due to proximity effects was applied. Such a one-dimensional model also cannot account for spatial coherence of ionizing background fluctuations on large scales. These limitations suggest that a one-dimensional model is insufficient to study the effect of opacity fluctuations on the \HeIIa ionizing background, motivating the three-dimensional approach described in the following section. 

In this work, we apply the \citet{FG2009} model for $\eta(N_\HI)$ to compute $d\bar{\tau}/dz$ as a function of the \emph{local} intensity of the \HeIIb-ionizing background. This procedure assumes that the \HeIIa fraction in the absorbers responds instantaneously to changes in $\GHeII$. In reality, the \HeIIa fraction will change on a characteristic timescale $t_\mathrm{eq} = (\GHeII+\alpha_\HeII n_e)^{-1} \sim 10 (\GHeII/10^{-14.5})^{-1}$ Myr. If the average quasar lifetime is shorter than this timescale then non-equilibrium ionization effects could be very important \citep{McQuinn2009b}. All other analytic treatments of the \HeIIb-ionizing background have made the same assumption with respect to the absorbers of \HeIIb-ionizing photons (e.g. \citealt{Fardal1998,FG2009,HM2012}).

\section{A Numerical Model of the He II Ionizing Background}\label{sec:methods}

In this work, we present a simple three-dimensional extension to the 1D model of DF14. The basic structure of the model is as follows. Quasars are randomly placed in a cosmological volume 500 Mpc on a side from $z=4$ to $z=2.5$ following the \citet{Hopkins2007} $B$-band quasar luminosity function (QLF). By placing quasars randomly we neglect their clustering; the effect of quasar clustering on fluctuations in the ionizing background is likely small \citep{Dixon2014} but could play a role when the mean free path is very short (\citealt{Desjacques2014}; see Section~\ref{sec:caveats}). The ionizing spectrum of each quasar is determined by first converting the $B$-band luminosity to the luminosity at the \HIa ionizing edge with the constant conversion factor from \citet{Hopkins2007} and assuming that the spectrum at $\nu>\nu_\HI$ is a power law $L_\nu \propto \nu^{-\alpha_\mathrm{Q}}$ with $\alpha_\mathrm{Q}=1.6$ in agreement with \citet{Telfer2002} and consistent with the recent estimate by \citet{Lusso2015}. We assume isotropic emission of ionizing radiation and a ``lightbulb" model for quasar light curves with a lifetime of 50 Myr, similar to \citet{Compostella2013,Compostella2014}. The ionizing background due to these quasars is then calculated at all points on a 50$^3$ grid, giving a spatial resolution of 10 Mpc. This resolution is sufficient because the fluctuations in the ionizing background that we hope to characterize will only manifest on scales larger than the typical mean free path inside the simulation volume ($\ga20$ Mpc), with the dominant scale set by the distance between bright sources ($\sim40$ Mpc, cf. DF14). Additionally, the simplified absorber physics from Section 2 is likely only a reasonable description on large-scales, so higher resolution would not add to our understanding.

The specific intensity of ionizing radiation $J_\nu$ at each point $\vec{r}$ on the grid is computed by adding together the contribution from every quasar $i$ with specific luminosity $L_i(\nu)$ at position $\vec{r}_i$ that is turned on at time $t - |\vec{r}-\vec{r}_i|/c$,
\begin{equation}
J_\nu(\vec{r},t) = \sum_{i} \frac{L_i(\nu)e^{-\tau(\vec{r},\vec{r}_i,\nu)}}{(4\pi|\vec{r}-\vec{r}_i|)^2},
\end{equation}
where the optical depth of ionizing radiation from the quasar $\tau(\vec{r},\vec{r}_i,\nu)$, which in previous work has been approximated as $\sim|\vec{r}-\vec{r}_i|/\lambda_\mathrm{mfp}$, is computed by integrating the fluctuating IGM opacity along the light cone,
\begin{equation}
\tau(\vec{r},\vec{r}_i,\nu) = \int_{\vec{r}_i}^{\vec{r}} \frac{d\tau}{dz} \frac{dz}{dl} dr',
\end{equation}
where $d\tau/dz$ is given by equation (1) evaluated at the redshift corresponding to $t' = t - |\vec{r}'-\vec{r}|/c$, and which depends on the local ionization rate $\GHeII(\vec{r}',t')$ through the absorber model discussed in Section 2. The ionization rate is then computed by integrating over frequency,
\begin{equation}
\GHeII(\vec{r},t) = 4\pi\int_{\nu_\HeII}^\infty \frac{J_\nu(\vec{r},t)}{h\nu} \sigma_\HeII(\nu) d\nu,
\end{equation}
where in practice this integral is computed discretely with 12 logarithmic bins in $\nu$ from $\nu_\HeII$ to $10^{1.2} \nu_\HeII$. We neglect the redshifting of photons as they travel from the quasar to the cell and use the proper distance to the source rather than the luminosity distance when computing $J_\nu$, but these should be relatively small effects. Because the opacity between the cell and the quasar depends on the \emph{history} of the opacity between the source and cell, a fully self-consistent model requires a time-dependent fluctuating background. We adopt a fiducial time step of 5 Myr, corresponding to $\sim6$--$7$ Mpc of light travel distance per step.

Because the IGM opacity depends on the local value of $\GHeII$, which in turn depends on the opacity, as discussed in Section~\ref{sec:opacity}, the calculation must be iterated in order to achieve self-consistency. We iterate until the average change of $\GHeII(\vec{r},t)$ across all grid cells between iterations is less than 1 per cent at $z=3$, which typically requires 7--8 iterations.

We have additionally run a ``control" simulation that ignores spatial fluctuations in the mean free path. Instead, $d\tau/dz$ is spatially uniform and computed as a function of redshift from a standard 1D ionizing background model with the same input parameters (i.e. quasar emissivity, CDDF). The rest of the computation is done in a similar manner to the full model described above, including finite quasar lifetimes and light cone effects. We will refer the this model as the ``uniform MFP" model in the rest of the paper.

\subsection{Results}

\begin{figure}
\begin{center}
\resizebox{8cm}{!}{\includegraphics{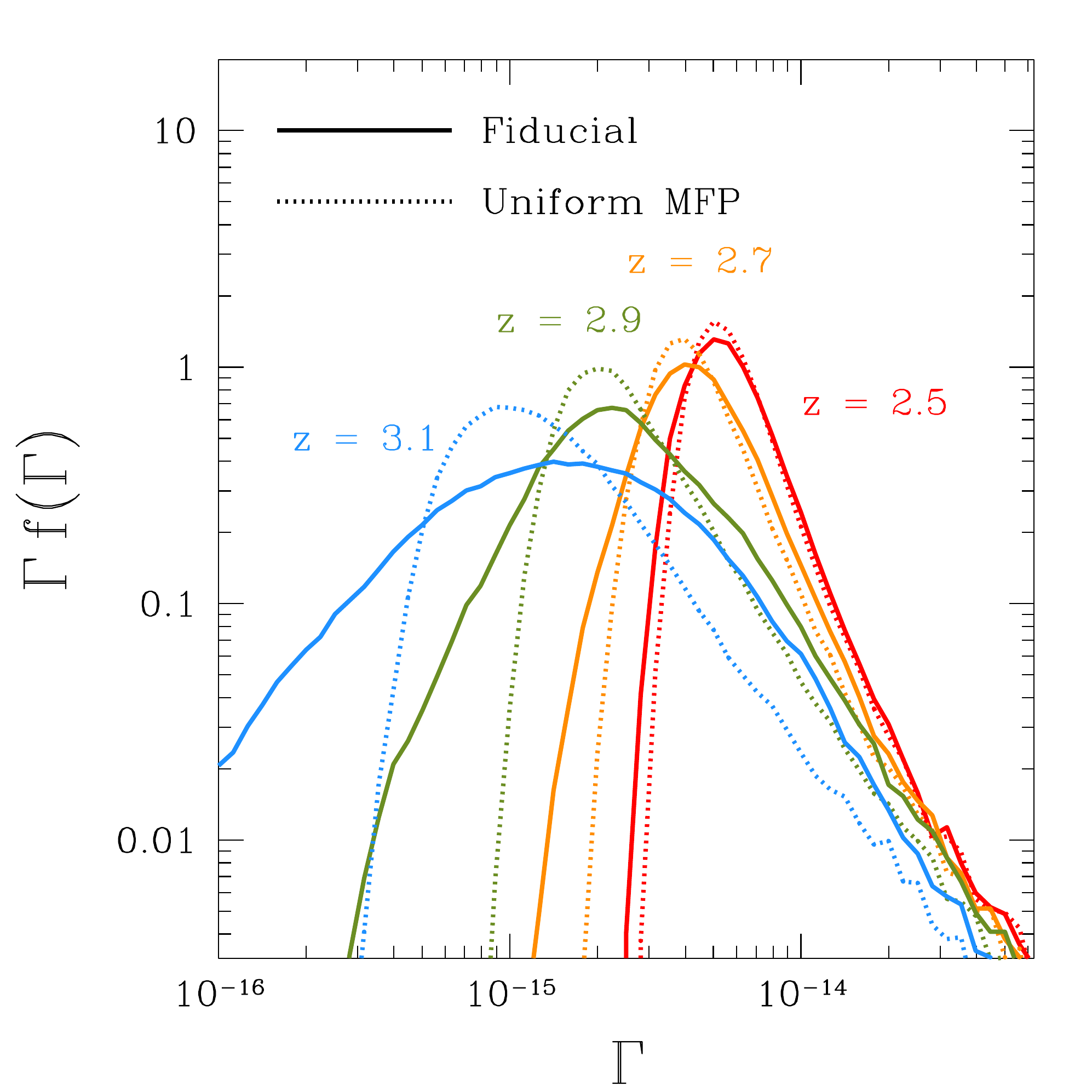}}\\
\end{center}
\caption{The solid curves show $\Gamma f(\Gamma)$ from the fiducial ionizing background simulation for $z=(2.5,2.7,2.9,3.1)$ from right to left. The dotted curves show the distributions from the uniform MFP model at the same redshifts. The differences between the two sets therefore demonstrate the effect of mean free path fluctuations.}
\label{fig:fgamma}
\end{figure}

\begin{figure*}
\begin{center}
\resizebox{15cm}{!}{\includegraphics{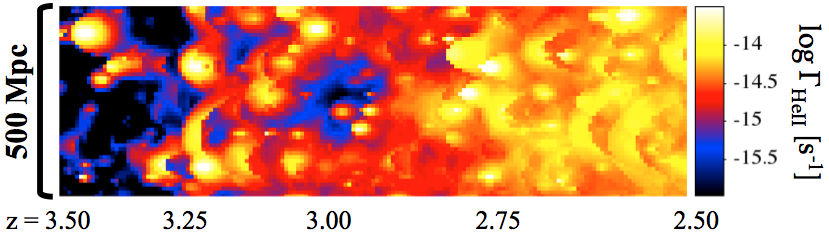}}\\
\end{center}
\caption{Light cone projection of $\Gamma_\HeII$ from $z=3.5$--$2.5$ in the fiducial fluctuating mean free path simulation. The vertical axis is position on the sky, and the horizontal axis is distance along the light cone with the observer located to the right. The parabolic features are due to the intersection of the light cone with a nearby bright quasar, though the exact geometry depends on the transverse distance. Large-scale correlations are seen along the light cone due to the coherence of ionization structures in the fluctuating mean free path model.
}
\label{fig:lightcone}
\end{figure*}

\begin{figure*}
\begin{center}
\resizebox{13cm}{!}{\includegraphics{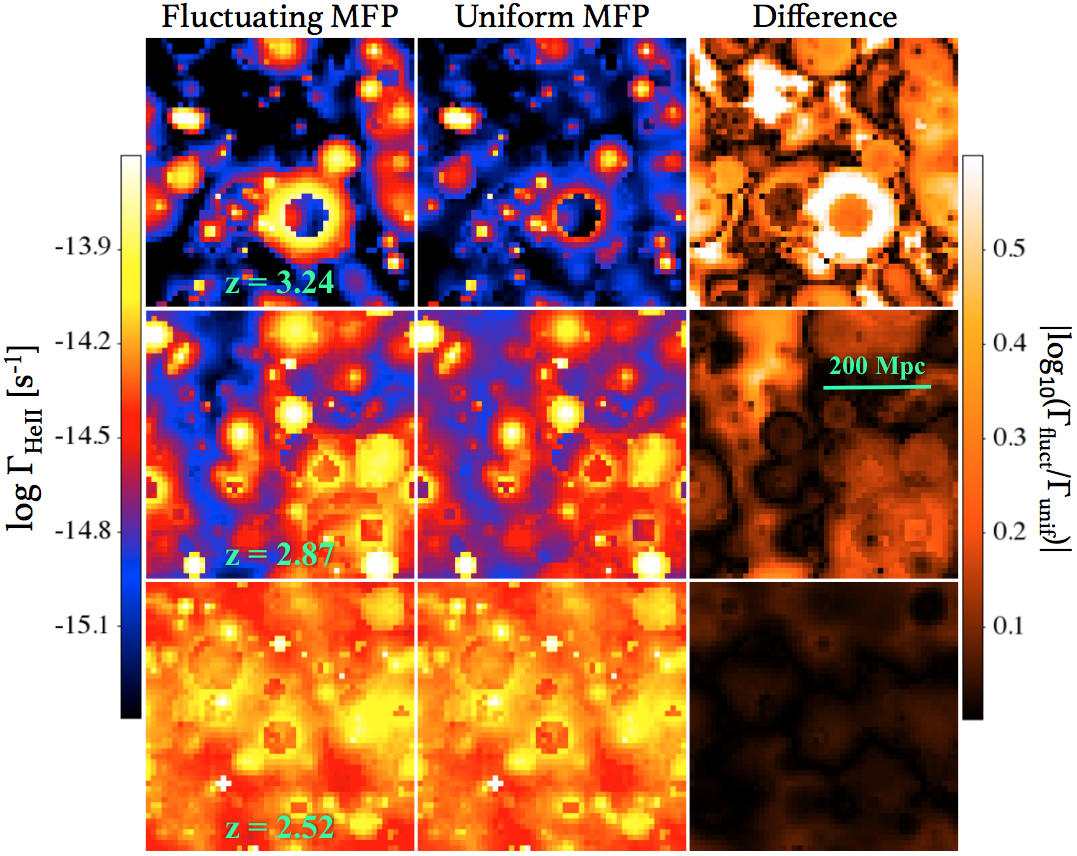}}\\
\end{center}
\caption{Slices of the simulation volume at $z=3.24$, 2.87, 2.52 from top to bottom. The left and middle panels show the \HeIIa ionization rate from the fluctuating and uniform mean free path simulations, respectively. The right panel shows the difference between the two, expressed as the ``distance" between the models in log space. The ionizing background becomes significantly more uniform over time and as a result the effect of including a fluctuating mean free path decreases. The topology of the fiducial model evolves from proximity-zone dominated at $z\ga3$ (top), to highly fluctuating at $2.7\la{z}\la3.0$ with large-scale correlations (middle), to mostly uniform at $z\la2.7$ (bottom).
}
\label{fig:snapshots}
\end{figure*}

The solid curves in Figure~\ref{fig:fgamma} show the evolution of the distribution of ionization rates, $f(\Gamma)$, from the simulation volume at $z=3.1$--$2.5$ in steps of $dz=0.2$. The dotted curves show the distributions for the uniform MFP model, which are nearly identical to analytic models for $f(\Gamma)$ computed using the same mean free path \citep{MW2003,Furlanetto2009}. The distributions in the full simulation are broader than the uniform MFP model due to the enhancement of large-scale features in the radiation field by the fluctuating mean free path, and the disagreement increases with increasing redshift as the mean free path fluctuations become more pronounced. In the fiducial model, at $z = $ (2.6, 2.8, 3.0, 3.2) the median and 16--84th percentile ionization rates are $\log{\GHeII} =$ (-14.33$^{+0.22}_{-0.13}$, -14.52$^{+0.29}_{-0.22}$, -14.75$^{+0.38}_{-0.35}$, -14.91$^{+0.47}_{-0.60}$), corresponding to $\lambda = $ (73$^{+25}_{-13}$, 45$^{+22}_{-12}$, 26$^{+19}_{-10}$, 17$^{+17}_{-10}$) Mpc.

The evolution of fluctuations in the helium-ionizing background is shown visually in Figure~\ref{fig:lightcone} as a light cone projection from $z=3.5$--$2.5$. The parabolic features are caused by the intersection of the light cone with the expanding 50 Mly-thick ``light shell" from a particularly bright quasar near the sightline (similar to Figure 5 in \citealt{Croft2004}). At high redshift, the mean background is dominated by the relatively transparent proximity zones around luminous quasars. By $z\sim2.5$, the mean free path is $\ga100$ Mpc, leading to modest background fluctuations of about a factor of two that are similar to observations \citep{MW2014}. Figure~\ref{fig:snapshots} shows a series of snapshots of a slice through the simulation volume, with $\Gamma$ from the uniform and fluctuating mean free path models and maps of the logarithmic difference between the two. The dominant features change from transparent proximity zones at $z>3$ (top row) to large-scale ($\sim200$ Mpc) coherent structures at $z\sim2.9$ (middle row) before finally the ionizing background becomes mostly uniform and unaffected by mean free path fluctuations at $z<2.7$ (bottom row). At early times, the fluctuating model has substantial differences from the uniform model principally because the mean free path is substantially larger near the bright sources, leading to larger proximity zones (or ``fossils" once the quasar has turned off), and deeper opacity far from those sources where the mean free path becomes very small.  At later times ($z=2.87$ in Figure~\ref{fig:snapshots}), the amplitude of these effects decreases, but their large-scale coherence remains, thanks to the rarity of the sources.

From Figure~\ref{fig:fgamma} we see that the high-$\Gamma$ end of $f(\Gamma)$ ($\log{\Gamma} \ga -13.8$) changes very little with redshift. These high-$\Gamma$ regions typically lie in the \emph{transparent} proximity zones of bright quasars (DF14), which do not evolve significantly because they are largely decoupled from the global ionizing background evolution (i.e. the \emph{local} mean free path is long enough that $\Gamma(r)\propto r^{-2}$ is a good approximation). The small variation with redshift that remains reflects fluctuations in the (small) number of extremely luminous quasars present in the simulation volume.

\begin{figure}
\begin{center}
\resizebox{8cm}{!}{\includegraphics{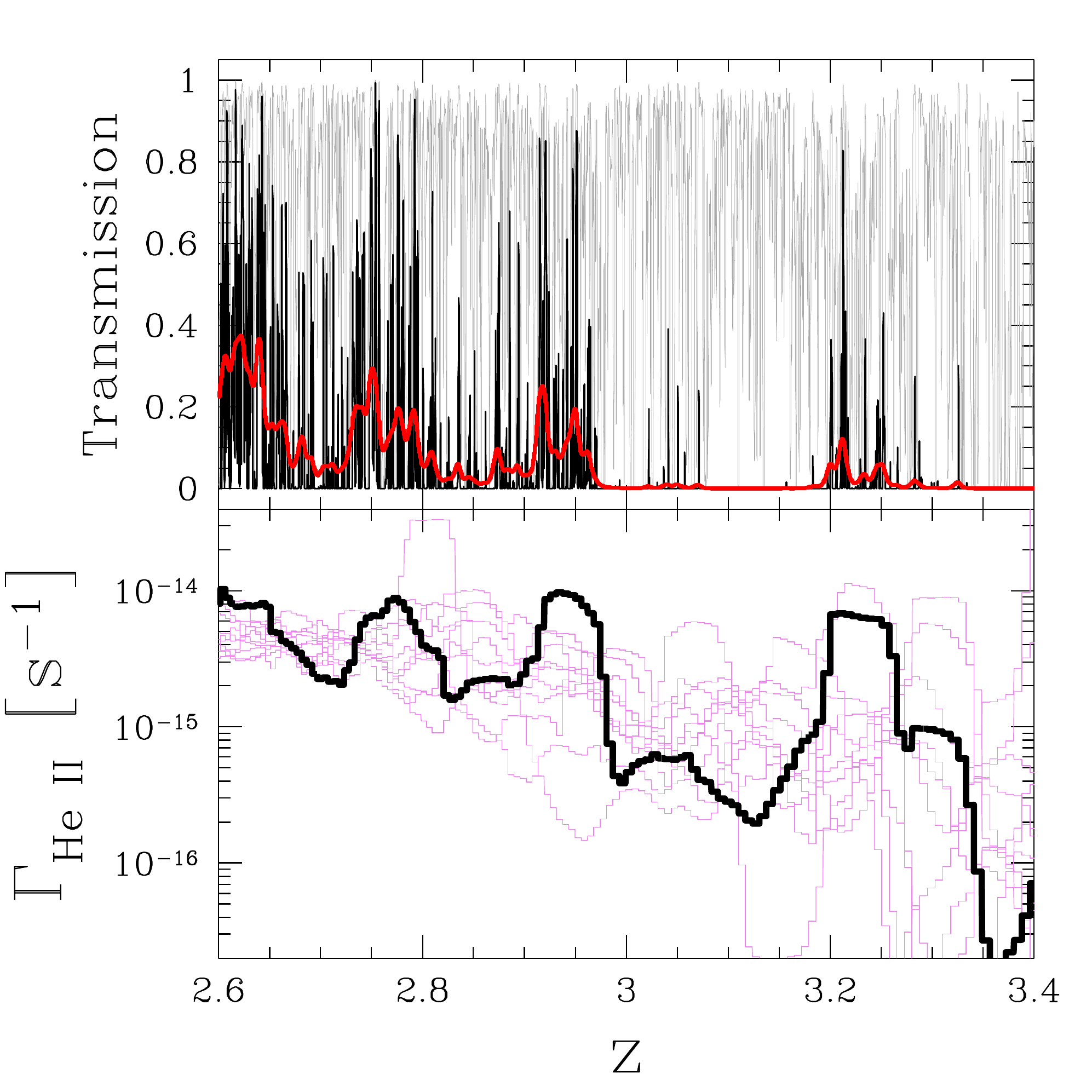}}\\
\end{center}
\caption{Top panel: Mock \HeIIa \lya forest (black) and \HIa forest (grey) transmission spectrum for a single light cone sightline. Both transmission spectra are shown at the arbitrarily high resolution of our simulations and do not include instrumental smoothing or noise. The red curve shows the \HeIIa transmission smoothed to mimic the resolution of HST/COS G140L. Bottom panel: $\Gamma_\HeII$ along the same sightline (thick curve) along with ten other random sightlines (thin curves).} 
\label{fig:mock}
\end{figure}

\begin{figure}
\begin{center}
\resizebox{8cm}{!}{\includegraphics{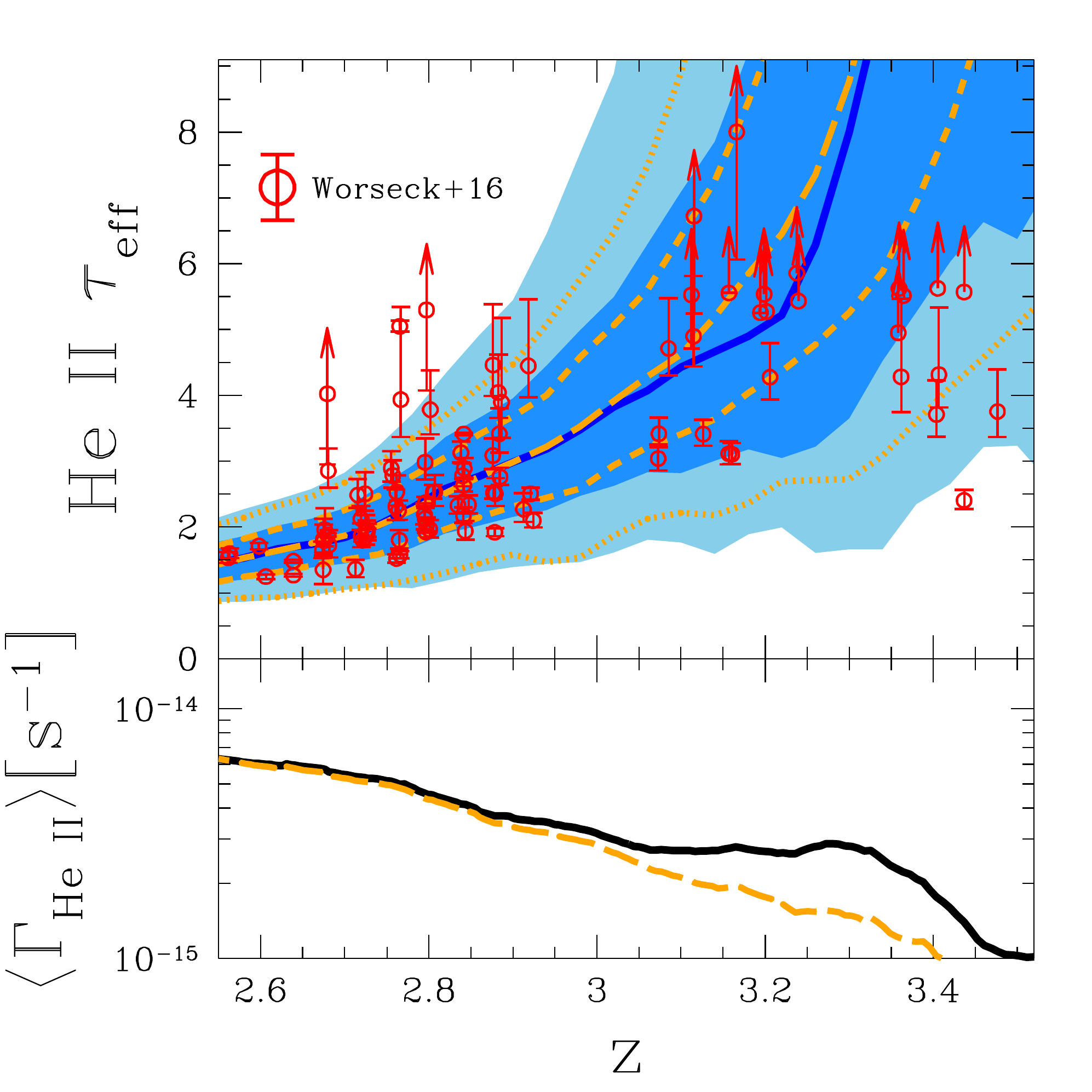}}\\
\end{center}
\caption{Top: Evolution of the \HeIIa effective optical depth in $\Delta z=0.04$ ($\approx 40$ Mpc) bins. The solid curve shows the median \HeIIa $\teff$ in our fiducial (500 Mpc)$^3$ ionizing background simulation, and the dark and light shaded regions show the 16--84th and 2.5--97.5th percentiles, respectively. Long-dashed, dashed, and dotted curves show the median optical depth and 16--84th and 2.5--97.5th percentiles for the uniform MFP model (see text) normalized slightly to match the fiducial simulation at $z\sim2.5$. The red data points are the $\Delta z=0.04$ measurements by \citet{Worseck2014a} with small $\Delta z$ offsets for clarity. Bottom: The solid curve shows the evolution of the mean $\GHeII$ in the simulation. The long-dashed curve is the mean $\GHeII$ in the uniform MFP model, increased by $\sim7$ per cent to match the fiducial model at $z\sim2.5$.}
\label{fig:opdepth}
\end{figure}

\section{Predictions for the He II Ly$\alpha$ Forest}\label{sec:forest}

The primary observable of \HeIIa is the opacity of the \HeIIa \lya forest ($\lambda_\mathrm{rest} < 304$ \AA) in far-ultraviolet spectra of $z\sim3$--$4$ quasars. In this section we describe a simple model of mock \HeIIa \lya transmission spectra through the fluctuating ionizing background computed in the previous section.

The foundation of our mock spectrum model is a Monte Carlo realization of Poisson-distributed absorbers drawn from the observed CDDF, computed in a manner similar to \citet{WP2011}. We draw \HIa absorbers in the range $10^{11.3} < N_\HI < 10^{19}$ cm$^{-2}$ from the same evolving CDDF used in Sections \ref{sec:opacity} and \ref{sec:methods} to compute the ionizing continuum opacity and place them randomly along a sightline. The \HeIIa content of each absorber is computed as in Section~\ref{sec:opacity}, assuming a constant uniform $\Gamma_\HI=10^{-12}$ s$^{-1}$ \citep{BB2013} and $\GHeII$ extracted along the light cone from the fluctuating \HeIIa ionizing background simulation. Doppler widths for each absorber are drawn in the range $10 < b < 100$ km/s from the distribution function $dN/db \propto b^{-5} \exp{[-b_\sigma^4/b^4]}$ \citep{HR1999} with $b_\sigma = 24$ km/s \citep{Kim2001}, and we assume that $b_\HI = b_\HeII$ (i.e. the Doppler widths are dominated by nonthermal motion). Finally, we compute transmission spectra assuming Voigt profiles for each absorber using the efficient approximation of \citet{TG2006}. 

An example \HeIIa (black) and \HIa (grey) transmission spectrum is shown in the top panel of Figure~\ref{fig:mock}. The transmission through the \HeIIa forest is closely tied to the variations in $\GHeII$ along the sightline, shown in the bottom panel. Large-scale regions along the light cone with enhanced $\GHeII$ are apparent, which share a characteristic width corresponding to our chosen quasar lifetime. Our \HeIIa \lya forest model has effectively infinite resolution and shows narrow regions with transmission close to unity, but more closely resembles observations when smoothed to mimic the resolution of HST/COS G140L grating, shown by the thick red curve.

We then binned mock spectra from 3000 randomly directed sightlines into $\Delta z=0.04$ pieces to allow direct comparison to the observations of \citet{Worseck2014a}. We compute the effective optical depth of each bin as $\tau_\mathrm{eff} = -\ln(\sum T_i/N)$ where $T_i$ is the transmission in each of the $N$ pixels inside the bin. The solid curve in Figure~\ref{fig:opdepth} shows the evolution of the median effective optical depth in the full simulation. The median optical depth increases steadily from $\tau_\mathrm{eff}\sim2$ at $z\sim2.7$ to $\tau_\mathrm{eff}\sim5$ by $z\sim3.1$, similar to the observations by \citet{Worseck2014a}. The dark and light shaded regions in Figure~\ref{fig:opdepth} show the 16--84th and 2.5-97.5th percentiles of $\tau_\mathrm{eff}$ values. Across the entire redshift range the low-$\tau$ end of the distribution, corresponding to regions with high $\GHeII$, evolves slowly. In contrast, the high-$\tau$ end, corresponding to regions with low $\GHeII$, increases rapidly above $z\sim2.8$. 

The thin dashed and dotted curves in Figure~\ref{fig:opdepth} show the median and distributions of $\teff$ in the uniform MFP simulation, re-normalized slightly to match the median optical depth in the fiducial model at $z=2.55$. The median $\teff$ evolution of the uniform MFP simulation is very similar to the full model, but the distribution tends to be more narrow. Fluctuations in the mean free path are most important in the tails of the distribution, as seen by the growing 2.5--97.5th percentile width at $z\ga2.8$. DF14 predicted that the evolution of $\Gamma$ should be accelerated by including fluctuations in the mean free path because the IGM opacity increases when averaged over $f(\Gamma)$. Instead, in our 3D model the evolution of $\Gamma$ from the uniform MFP model is strikingly similar to the fluctuating MFP model at $z\la3$. This lack of excess opacity from fluctuations is likely due to the fact that ionizing photons are preferentially emitted from bright quasars with transparent proximity zones, so the distribution of $\Gamma$ seen by ionizing photons is biased towards higher values.

The mock distribution is similar to the observations in \citet{Worseck2014a} (red points), with the exception of a handful of high-$\teff$ regions at $z\sim2.75$ and a very low $\teff$ region at $z=3.44$. In particular, the well-measured $\teff\sim5$ region seen towards HE2347-4342 is quite rare in our simulations, appearing only in $\sim0.1$ per cent of sightlines. There is some evidence for disagreement at $z\sim3.4$ -- our model predicts that only $\sim1$--$2/10$ $\Delta z=0.04$ segments should show detectable transmission (i.e. $\teff\la5.5$) while the observed fraction is $5/10$. We leave a full statistical comparison of our modeled optical depth distributions to observations, including forward modeling of instrumental noise, to future work.

The upper and lower bounds to the $\teff$ distribution are driven by the volume of space far away from (i.e. $\tau(\vec{r},\vec{r}_\mathrm{Q},\nu_\HeII) > 1$) and very close to luminous quasars, respectively. The former evolves very quickly with redshift -- not only is the number density of luminous quasars decreasing above $z\sim2.5$, but more importantly the mean free path also decreases very rapidly. Additionally, as one moves further away from luminous sources in real space, the weaker ionizing background causes the mean free path to shrink, causing a non-linear increase in the optical depth. The lower bound to the $\teff$ distribution evolves only due to the evolution of the bright end of the QLF, with some fluctuations due to Poisson variance in the number of very luminous quasars -- the ``bumpy" evolution in our model is due to this variance rather than poor sampling of the $\teff$ distribution.

Our model allows a considerable amount of ``tuning" of parameters to reproduce the smoothly evolving median $\teff$ at $z\la3$, namely through the emissivity of \HeIIb-ionizing photons (via the quasar spectral index $\alpha_\mathrm{Q}$), the ionization state of \HIa absorbers (via $\GHI$), and the exact form of the CDDF (via the shape, normalization, and minimum $N_\HI$). It is actually a remarkable coincidence that -- without deliberate tuning -- our fiducial set of model parameters provides good agreement to the \HeIIa $\teff$. We discuss the effect of variations in model parameters further in Section~\ref{sec:uncertain}. 

\begin{figure}
\begin{center}
\resizebox{8cm}{!}{\includegraphics{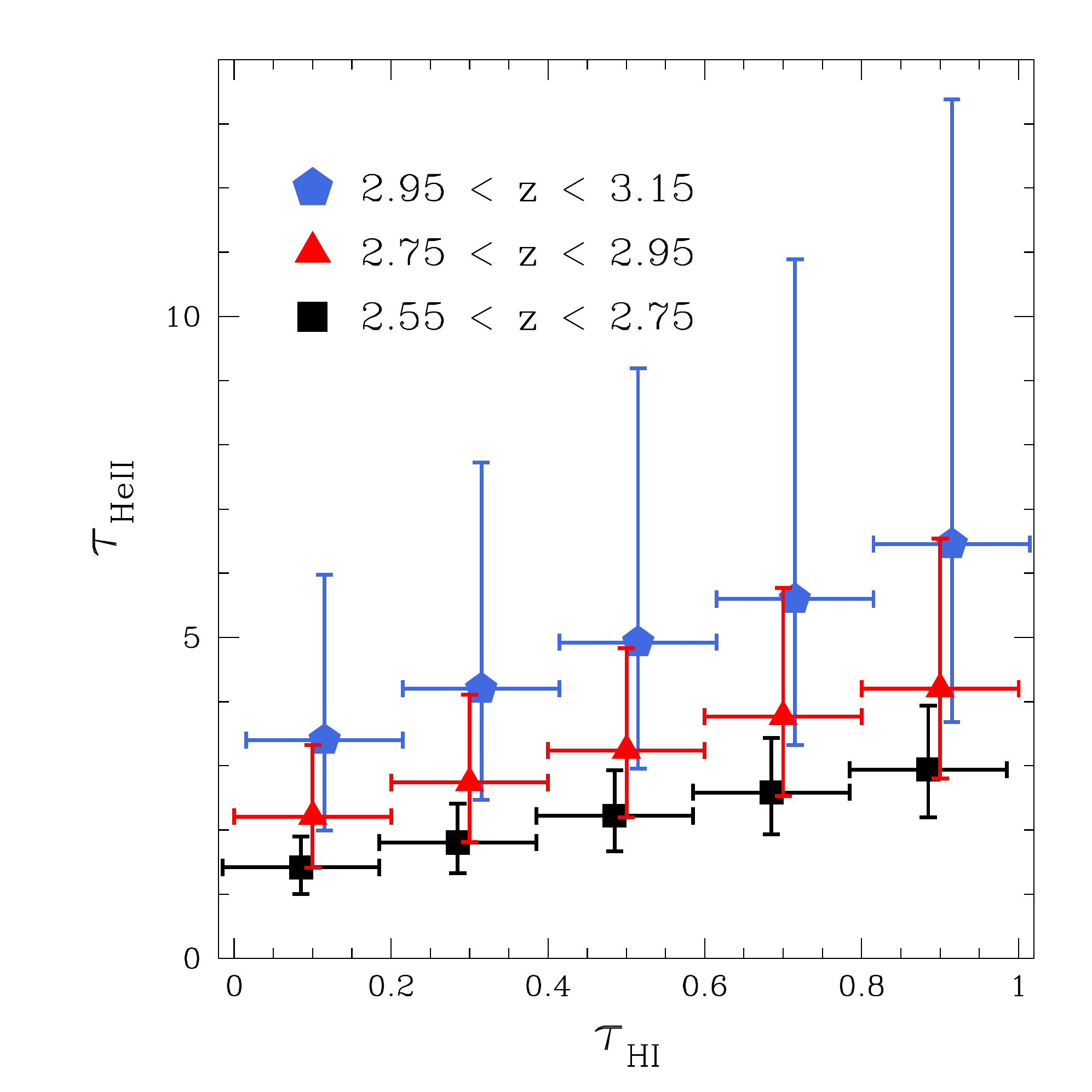}}\\
\end{center}
\caption{Distribution of \HeIIa optical depths at $\Delta z=0.01$ scales, roughly corresponding to the spatial resolution of our ionizing background model, as a function of \HIa optical depth. The black squares, red triangles, and blue pentagons show the median $\tau_\HeII$ in bins of $\Delta\tau_\HI=0.2$ for $z\sim2.65$, $2.85$, and 3.05, respectively. Error bars show the 16--84th percentiles of $\tau_\HeII$ within each bin. The points and \HIa ranges have been shifted slightly between the redshift ranges for clarity. Compare to Figure 8 of \citet{Worseck2014a}.
}
\label{fig:tauhr}
\end{figure}

In Figure~\ref{fig:tauhr} we compare the associated \HIa and \HeIIa $\teff$ on scales of $\Delta z = 0.01$ ($\approx 10$ Mpc) for comparison to Figure 8 of \citet{Worseck2014a}. This scale is also coincidentally the spatial resolution of our simulations, and thus the smallest scale that we can make predictions for variations in $\teff$ due to the fluctuating ionizing background. The behavior of the median optical depth is very similar to that of the \citet{Worseck2014a} data at all redshifts. However, at $z\sim2.85$ our simulations fail to reproduce the substantial observed fraction of regions with $\tau_\HI\la0.5$ and $\tau_\HeII\ga6$, indicating that the Universe has stronger (rare) downward fluctuations in $\GHeII$ at this epoch than our simulations produce. In addition, one might expect \emph{weaker} fluctuations due to the correlation between absorber systems and quasars, which we explicitly ignore and likely plays a stronger role in the distribution of effective optical depths at these scales compared to $\Delta z = 0.04$. Nevertheless, the agreement of our median relationship between \HIa and \HeIIa suggests that our modeling approach is not unreasonable. It is worth noting that for $z\ga3$ a large fraction of sightlines have little-to-no detectable flux given current observational limitations \citep{Worseck2014a}, so judging the agreement between our simulation and observations is difficult.

\section{Model Uncertainties}\label{sec:uncertain}

We have shown that a relatively simple model of fluctuations in the helium ionizing background can reproduce most of the observed properties of the \HeIIa \lya forest across cosmic time. This implies that fluctuations in the radiation field should be considered a viable alternative to ongoing \HeIIa reionization (e.g. \citealt{Worseck2011}) as an explanation of the observed $\teff$ variations, at least for $z\la3.2$. We discuss some important caveats to our simple model assumptions below.

\subsection{Caveats}\label{sec:caveats}

\noindent\emph{(i) \HeIIa \lya forest prescription}

Our model assumes that the \HeIIa \lya forest can be described as an ensemble of randomly distributed absorbers following the \citet{Prochaska2014} CDDF. While this kind of description has been successfully used to model the \HIa effective optical depth in previous work (e.g. \citealt{HM2012}), \HeIIa effective optical depths greater than $\sim2.0$ at $z\la3$ require near-complete blanketing of the spectrum by numerous low $N_\HI$ ($\la 10^{12}$ cm$^{-2}$) lines \citep{Fardal1998}. The abundance of such ``absorbers" is completely unconstrained by observations, and physical interpretation via the Jeans ansatz \citep{Schaye2001} becomes unrealistic with implied absorber sizes of $\ga 2$ Mpc. In effect, these lines act as a ``smooth" component to the \HeIIa \lya absorption that is required to reproduce the observed $\teff\ga1.5$ absorption. Our treatment of the \HeIIa \lya forest as a set of discrete absorption lines is a rough approximation, with the low $N_\HI$ systems representing the smooth low density environments of cosmic voids that provide the majority of the \HeIIa \lya opacity \citep{Croft1997}. 

To judge whether our \HeIIa \lya forest model responds to a fluctuating ionizing background similar to a more realistic density field (at least on large scales), we performed a simple comparison to skewers through a \textsc{\small NYX} Eulerian hydrodynamical simulation \citep{Almgren2013,Lukic2015}, kindly provided to us by Z. Luki{\'c}. The simulation is 100 $h^{-1}$ Mpc on a side with 4096$^3$ gas elements and dark matter particles, and we use a single output of the density and velocity fields at $z=3$ for our comparison. To calculate \HeIIa \lya transmission spectra of the skewers, we computed the equilibrium \HeIIa fraction of each cell assuming a range of uniform $\GHeII$ and included the effects of thermal broadening and redshift-space distortions due to peculiar velocities. We found that the normalization, scaling of $\teff$ with $\GHeII$, and variations in $\GHeII$ on scales of $\Delta z=0.04$ was very similar to our model over the range of relevant values. In detail, the median $\teff(\GHeII)$ relationship in the hydrodynamical simulation was slightly steeper, $\tau_\mathrm{eff,hydro}\propto\GHeII^{-0.54}$ vs. $\tau_\mathrm{eff,CDDF}\propto\GHeII^{-0.47}$, with a $\sim10$ per cent lower normalization. These differences are not large enough to significantly change our results.

\noindent\emph{(ii) Ionization equilibrium}

Our simulations assume that the universe is in ionization equilibrium -- that is, reionization is assumed to have completed some time in the past. Assuming the gas density probability distribution from \citet{ME2000} and following their procedure to compute the neutral fraction of the IGM, the equilibrium \HeIIa fraction in our simulation is below $1$ per cent by volume at $z\la3$. For the lowest $\GHeII$ regions at $z\sim3$ the equilibrium \HeIIa fraction can be as high as $\sim10$ per cent, so an equilibrium treatment is unlikely to be very accurate. This is the primary weakness of our model compared to full radiative transfer simulations (e.g. \citealt{McQuinn2009,Compostella2013,Compostella2014}).

\noindent\emph{(iii) Clustering of sources and absorbers}

In both the source and absorber models we neglect the effect of clustering. As mentioned previously, \citet{Dixon2014} showed that the impact of quasar clustering on ionizing background fluctuations was likely small. Recent work by \citet{Desjacques2014} suggests that clustering could have a significant effect on background fluctuations at $z\sim3$ if the mean free path is comparable to the correlation length of quasars, $r_\xi\sim15$ Mpc. The average mean free path in our simulations is substantially larger than this for the redshifts we are interested in ($\lambda_\mathrm{mfp}\ga25$ Mpc), but it is reasonable to expect that the addition of mean free path fluctuations would enhance the effect of clustering to some degree. 

We ran an additional ionizing background simulation with clustered sources to test this possibility. Using the semi-numerical simulation code \textsc{\small DEXM} \citep{MF2007}, we generated a realistic distribution of dark matter halos in a (500 Mpc)$^3$ volume at $z=3$. We then re-ran one of our ionizing background simulations with the quasars randomly assigned to locations of the most massive $N_\mathrm{QSO}$ halos, corresponding to $M_\mathrm{h}\ga1.3\times10^{12}$ M$_\odot$, and approximating the halo distribution as constant with redshift. The median and low-$\teff$ end of the resulting opacity distribution were nearly identical to the fiducial, uniformly-distributed source model. At $z\la3$, the upper-$\teff$ end was boosted by $\sim10$ per cent and $\sim30$ per cent at the 84th and 97.5th percentiles, respectively, suggesting that the clustering of sources can increase the probability of high-$\teff$ outliers. While the model with clustering more closely reproduces the observed distribution of $\teff$ at $z<3$, detailed modeling of the highly uncertain connection between quasars and dark matter halos is outside of the scope of this work.

In constructing our mock \lya forest spectra the only large-scale fluctuations in the density field we consider are the Poisson variations in the number of absorber systems, which are expected to be fairly small \citep{Fardal1998}. Cosmic variance in the large-scale density field at the $\Delta z = 0.04 \approx 40$ Mpc scale considered in this work should also be small compared to the fluctuations in the ionizing background, with $\sigma(R=40\,\mathrm{Mpc})\sim0.1$. While the overall effect of \lya forest clustering may be small, its largest effect would likely be to extend the tails of the $\teff$ distribution, which could also ease some tension with the highest $\teff$ regions at $z\sim2.8$ in addition to including source clustering as described above. The correlations between the density field and the source field (i.e. the positions of quasars) could in principle lead to smaller fluctuations in $\teff$, but this is unlikely to be a strong effect -- the typical excess overdensity associated with a $\sim10^{12-13}$ M$_\odot$ dark matter halo on scales comparable to the $\sim40$ Mpc observed \HeIIa \lya forest bins should be $< 10$ per cent (see, e.g., \citealt{FG2008}).

\begin{figure}
\begin{center}
\resizebox{8cm}{!}{\includegraphics{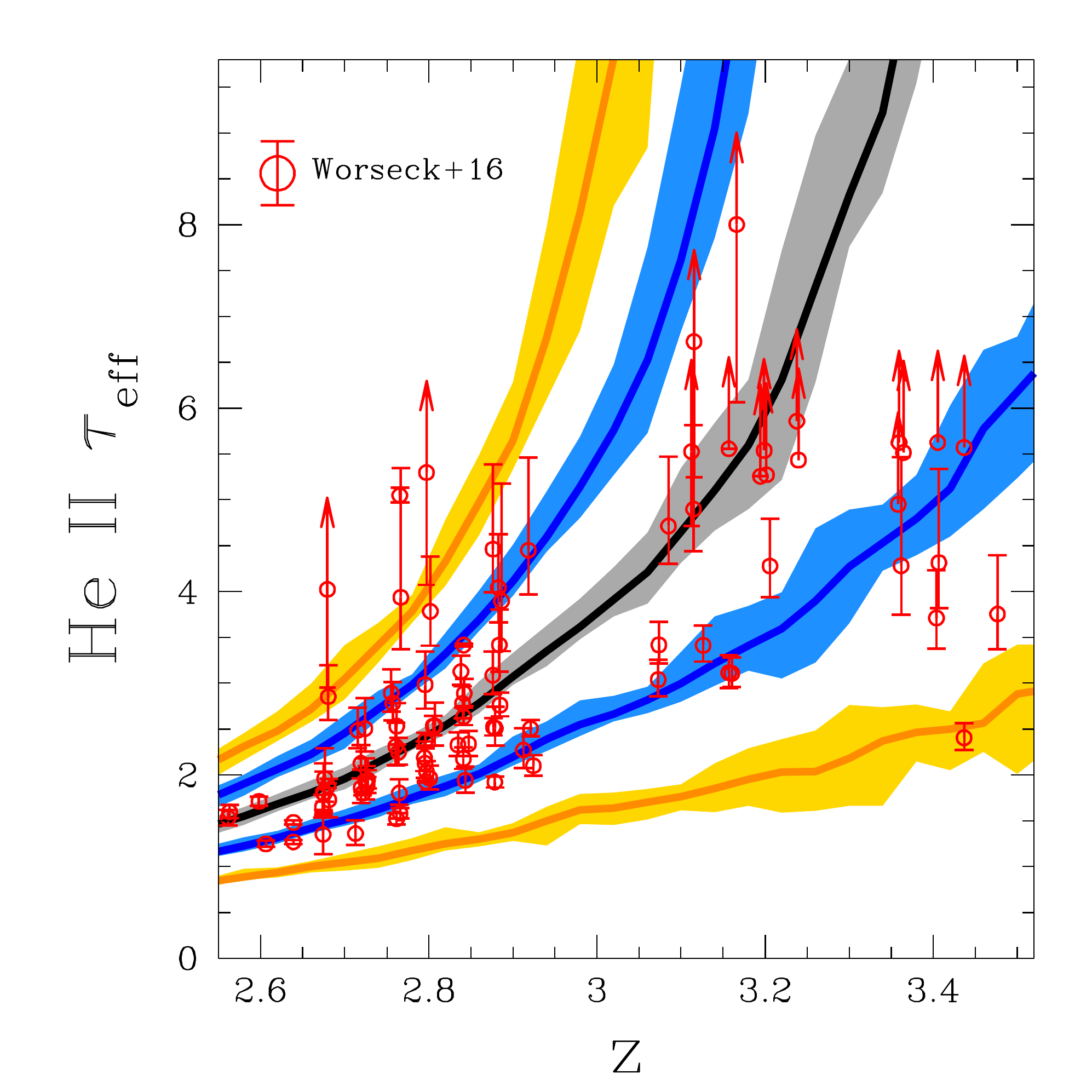}}\\
\end{center}
\caption{Cosmic variance of the \HeIIa $\teff$ distribution between nine different realizations of the (500 Mpc)$^3$ ionizing background model. The solid curves show the mean value of the median (black), 16--84th percentiles (blue), and 2.5--97.5th percentiles (orange). The corresponding shaded regions show the maximum and minimum values in the set of nine realizations. The \HeIIa measurements from \citet{Worseck2014a} are shown as the red open circles.}
\label{fig:cv}
\end{figure}

\subsection{Cosmic Variance}

Despite the large volume probed by our simulations, the mean optical depth and fluctuations that we measure are still affected somewhat by cosmic variance. The non-linear response of the ionizing background due to mean free path fluctuations implies that a random upward boost in the number of bright quasars leads to large-scale regions with larger-than-average mean free path that persist until the radiation from those quasars leaves the simulation volume. This can be seen in our fiducial simulation which has a pair of $\sim70 L_*$ quasars turn on at $z\sim3.4$. The ``boost" in $\GHeII$ and scatter of optical depths to low values from $z\sim3.2$--$3.4$ seen in Figure~\ref{fig:opdepth} is almost entirely attributable to these two sources. This demonstrates the sensitivity of our model to the number of the brightest sources, at least at early times when the average mean free path is small. 

In Figure~\ref{fig:cv} we show the $\teff$ distributions for a set of nine realizations of the (500 Mpc)$^3$ ionizing background model (including the fiducial one). The solid curves show the mean value of the median, 16--84th percentiles, and 2.5--97.5th percentiles, while the shaded regions correspond to the minimum and maximum values in the full set of realizations. Below $z\sim3$ the $\teff$ distributions between realizations agree to within $\pm10$ per cent, but at higher redshift the range increases to $\ga20$ per cent. This suggests that simulations which hope to accurately predict the post-reionization $\teff$ distribution should be at least the size of our model (i.e. $\ga500$ Mpc on a side), but even larger volumes may be required for convergence at the $10$ per cent level. Achieving accurate radiative transfer in such large volumes will require substantial computational resources, far more than our simplified approach.

\subsection{Variations in Model Parameters}\label{sec:params}

The primary input parameters to our model are: the quasar luminosity function, the quasar ionizing spectral index $\alpha_\mathrm{Q}$, the quasar lifetime $t_\mathrm{Q}$, the strength of the \HIb-ionizing background $\GHI$, and the column density distribution of \HIa absorbers $f(N_\HI,z)$. We leave a full analysis of variations in these parameters to future work, but we qualitatively describe their effects and importance below.

We have applied the \citet{Hopkins2007} QLF to populate the quasars in our simulation. More recent measurements by \citet{Glikman2011} suggest that there are substantially more faint quasars than the \citet{Hopkins2007} QLF at $z\sim4$, which would result in a more uniform ionizing background. These measurements appear to be in conflict with the $z\sim4$ measurements by \citet{Masters2012}, but they instead find evidence for a somewhat steeper faint-end slope at $z\sim3$ (see also \citealt{Ross2013}). Accurate model predictions for the helium-ionizing background will require resolution of this discrepancy, which we do not attempt in this work. 

The primary effect of adjusting $\alpha_\mathrm{Q}$ is on the normalization of the quasar ionizing output at $\nu_\HeII$, because $L_\HeII = 4^{-\alpha_\mathrm{Q}} L_\HI$, where $L_\HeII$ and $L_\HI$ are the specific quasar luminosities at $\nu_\HI$ and $\nu_\HeII$, respectively. At fixed redshift, due to the self-consistency of absorbers with the ionizing background, the average ionization rate should roughly follow $\GHeII \propto \epsilon^{1/(2-\beta)}$ \citep{McQuinn2011}, where $\beta$ is the power law index of the CDDF in absorbers with \HeIIa ionizing opacity $\tau_{\nu,\HeII}\la1$ ($N_\HI\la10^{15.5}$ cm$^{-2}$). In the \citet{Prochaska2014} CDDF model that we adopt, $\beta = 1.6$--$1.75$ in the relevant range of column densities, so we expect $\GHeII \propto \epsilon^{2.5\mathrm{-}4.0} \propto 4^{-(2.5\mathrm{-}4.0)\alpha_\mathrm{Q}}$. In practice, we find that $\GHeII \propto \epsilon^{2.5}$, so an adjustment of $\alpha_\mathrm{Q}$ by $0.1$ results in a change of $\GHeII$ by a factor of $\sim1.4$. This sensitivity is completely degenerate with any other adjustment in the ionizing emissivity, such as the conflict between the typical conversion of the \citet{Hopkins2007} rest-frame optical QLF and the UV luminosity density measured by \citet{Cowie2009}. Changing the shape of the \HeIIb-ionizing spectrum alone leads to a minor adjustment of the relationship between $\GHeII$ and the mean free path of average-energy ionizing photons that regulates the fluctuations in the background, so a harder spectrum would result in weaker fluctuations and a softer spectrum would lead to stronger fluctuations at fixed $\langle\GHeII\rangle$.

In our model we assume that quasars emit ionizing photons isotropically at a fixed rate for a fixed lifetime of 50 Myr and fixed spectral index $\alpha_\mathrm{Q}$. This is an enormously simplified picture of quasars that ignores realistic light curves (e.g. \citealt{HH2009}), anisotropic/beamed emission, and variations in the ionizing spectral index \citep{VB2001,Telfer2002}. We tested simulations with quasar lifetimes of 25 and 100 Myr and found that the main results of this work, the distribution of $\teff$ over $dz=0.04$, were qualitatively unchanged. The primary difference was in the characteristic width of features in $\GHeII$ along the light cone, as seen in Figures \ref{fig:lightcone} and \ref{fig:mock}. If quasar lifetimes are significantly shorter than the light travel time corresponding to $dz=0.04$, the blending of features would likely reduce the amount of fluctuations observed. However, on those short time scales ($\Delta{t} \la 10$ Myr), the equilibration time of the gas (Section~\ref{sec:opacity}) would become comparable to the quasar lifetime. 

In our model, $\GHI$ is fixed at a constant value, consistent with measurements by Becker \& Bolton (2013). Because $\lambda_\mathrm{mfp} \propto \eta^{-1} \propto \GHI$, adjusting $\GHI$ is roughly equivalent to changing the normalization of the mean free path at fixed $\GHeII$. In effect, we find that $\GHeII \propto \GHI^{-1}$, so for a given absorber in our \HeIIa \lya forest model, $N_\HeII \propto \GHI^{-2}$, and so very roughly $\teff \propto \GHI^{-2}$. If one assumed a smaller $\GHI$ (e.g. \citealt{FG2009}), a simple way to maintain the same $\teff$ would be to increase the \HeIIb-ionizing emissivity such that $\eta$ remained constant, or in other words, $\Delta \log{\GHeII} \sim 2\times \Delta \log{\GHI}$.

The redshift evolution of the mean free path and the details of its dependence on the local background are sensitive to the evolution and shape of the CDDF. The CDDF of \HIa absorbers has been well-measured at $z\sim2.5$, but discrepancies between different works exist in the difficult (i.e. saturated) $N_\HI$ regime that is most important to cosmological \HeIIa radiative transfer ($N_\HI\sim10^{15}$--$10^{16}$ cm$^{-2}$; see, e.g., \citealt{Kim2013,Rudie2013,Prochaska2014}). While recent observations have greatly increased our knowledge of the CDDF at $z\sim2.5$, the \emph{evolution} of both the shape and normalization of the CDDF are still very uncertain. The effects of different CDDF parameterizations on the evolution of the \HeIIb-ionizing background are discussed in detail in DF14.

Given the number of uncertain parameters in modeling both the sources and absorbers, there are important degeneracies in our model.  For example, for a fixed CDDF and model of the physical nature of absorbers, the measured average \HeIIa $\teff$ corresponds to a locus of reasonable combinations of the \HeIIb-ionizing emissivity and $\GHI$. In that sense, the matching of our fiducial model to the $z\sim2.5$ measurements of the \HeIIa $\teff$ is something of a coincidence -- other ``solutions" exist in a reasonable range of parameter space.  However, our results show that a self-consistent model of the \HeIIb-ionizing background, with parameters consistent with other measurements, is compatible with the vast majority of the forest observations. This conclusion does not result from a fortuitous choice of parameters but is true for a large swath of - though by no means the entirety of - parameter space.

\section{Conclusion}\label{sec:conclusion}

Recent observations of excursions to high effective optical depths in the \HeIIa \lya forest at $z \la 3$ have been interpreted as evidence for ongoing \HeIIa reionization \citep{Shull2010,Worseck2011}. We have shown that the majority of the scatter in $\teff$ measurements at $z \la 3.2$ can be explained by fluctuations in the ionizing background when mean free path fluctuations are included self-consistently. Our model consists of a 3D realization of randomly distributed quasars following the measured quasar luminosity function, with finite quasar lifetimes and a finite speed of light. We additionally let the mean free path vary depending on the strength of the local ionizing background in a manner analogous to standard 1D cosmological radiative transfer models \citep{FG2009,HM2012} that assumes ionization equilibrium throughout the IGM. 

The resulting radiation field fluctuates strongly on large scales, leading to large variations between sightlines. These large-scale features are due to the additional coherence caused by mean free path fluctuations, which are present even though we have neglected the clustering of sources in our fiducial simulations. If bright sources randomly lie close to each other, the excess background is enhanced due to the locally transparent IGM. Similarly, regions that are far from bright sources suffer from a more opaque IGM, further decreasing the radiation they receive. The strength of these effects increase strongly with redshift, such that by $z\ga3.2$ they dominate the structure of the ionizing background (see the top panels of Figure~\ref{fig:snapshots}). At these relatively high redshifts it is likely that the progression of \HeIIa reionization is important to the budget of \HeIIb-ionizing photons and the size of proximity zones, but it is currently unclear how to distinguish the two scenarios observationally. Despite the enhanced fluctuations of our model relative to a uniform ionizing background, a handful of observed regions with high optical depth at $z\sim2.7$--2.9 are in modest tension with our model predictions, suggesting that we may still be missing an important piece of the puzzle.

Recently, significant transmission in the \HeIIa \lya forest has been observed to persist to $z\sim3.5$ \citep{Worseck2014a}, which appears to be in (model-dependent) tension with late ($z\la3$) \HeIIa reionization \citep{McQuinn2009,Compostella2013,Compostella2014}. At face value, our model appears inconsistent with these observations because the likelihood of encountering large-scale $\teff \la 4$ regions is small. At these higher redshifts \HeIIa reionization is likely still ongoing, so our assumption of ionization equilibrium is no longer valid -- however, it seems likely that this would limit the size of ionized regions and exacerbate the tension. Resolution of this discrepancy could come from the addition of clustering to the quasar distribution (although our simple test of this scenario suggests otherwise) or changes in the evolution and shape of the QLF.

The large-scale variations in the \HeIIb-ionizing background we have described here show that \HeIIa reionization is a rich and complex event. The interaction of the sources, IGM, and radiation field require careful modeling, which is made possible by our detailed understanding of the IGM at $z \sim 3$. Observers are discovering more and more lines of sight along which the \HeIIa \lya forest can be studied, and the large fluctuations in our model imply that a full understanding of the reionization event will require exploring these new lines of sight in detail. 

\section*{Acknowledgements}
We would like to thank G. Worseck for valuable discussions, and for sharing his \HeIIa effective optical depth data with us prior to publication. We would also like to thank Z. Luki\'{c} for providing us with \textsc{\small NYX} hydrodynamical simulation output. SRF was partially supported by NASA grant NNX15AK80G, administered through the ATP program, and by a Simons Fellowship in Theoretical Physics. SRF also thanks the Observatories of the Carnegie Institute of Washington for hospitality while this work was completed. Some of these simulations were run on computers in the UK (Apollo at the University of Sussex). KLD was supported by the Science and Technology Facilities Council [grant number STI/L000652/1].

\end{document}